# Quantum-Efficient Convolution through Sparse Matrix Encoding and Low-Depth Inner Product Circuits

Mohammad Rasoul Roshanshah[1] . Payman Kazemikhah[1] . Hossein Aghababa[1,2]

**Abstract**
Convolution operations are foundational to classical image processing and modern deep learning architectures, yet their extension into the quantum domain has remained algorithmically and physically costly due to inefficient data encoding and prohibitive circuit complexity. In this work, we present a resource-efficient quantum algorithm that reformulates the convolution product as a structured matrix multiplication via a novel sparse reshaping formalism. Leveraging the observation that localized convolutions can be encoded as doubly block-Toeplitz matrix multiplications, we construct a quantum framework wherein sparse input patches are prepared using optimized key-value QRAM state encoding, while convolutional filters are represented as quantum states in superposition. The convolution outputs are computed through inner product estimation using a low-depth SWAP test circuit, which yields probabilistic amplitude information with reduced sampling overhead. Our architecture supports batched convolution across multiple filters using a generalized SWAP circuit. Compared to prior quantum convolutional approaches, our method eliminates redundant preparation costs, scales logarithmically with input size under sparsity, and enables direct integration into hybrid quantum-classical machine learning pipelines. This work provides a scalable and physically realizable pathway toward quantum-enhanced feature extraction, opening up new possibilities for quantum convolutional neural networks and data-driven quantum inference.

# 1 Introduction

Quantum algorithms [1–3] have demonstrated remarkable potential for accelerating

✉ Hossein Aghababa
  aghababa@ut.ac.ir

1 School of Electrical and Computer Engineering, Faculty of Engineering, University of Tehran, Tehran, Iran
2 Faculty of Engineering, College of Farabi, University of Tehran, Tehran,Iran.

computational tasks that are intractable for classical systems, most notably through landmark contributions such as Shor's integer factorization [1, 11], and Grover's search algorithm [6–8]. Building on these foundational results, the focus of quantum algorithm design has expanded into high-dimensional tensor operations, matrix arithmetic [9, 10], and data-driven inference [11, 12]. These areas are of particular importance due to the emergence of large-scale data applications, where classical methods face growing challenges related to memory bottlenecks and processing latency. Quantum computing offers a new paradigm for these problems by leveraging superposition, entanglement, and amplitude amplification to achieve improved scaling in both storage and computation.

Despite these advances, the design of quantum algorithms remains constrained by a narrow set of paradigms—primarily phase estimation [13], amplitude estimation [14], and Hamiltonian simulation. While these techniques have driven progress in quantum linear systems, quantum search, and adiabatic optimization [15], the field still lacks generalized frameworks for transforming practical, data-intensive operations into quantum-native procedures. In particular, matrix multiplication and convolution—two fundamental primitives in scientific computing and machine learning—remain challenging to implement efficiently due to the cost of state preparation and the difficulty of encoding large datasets into quantum states.

Convolution, in particular, plays a central role in signal processing, computer vision, and deep learning, especially within convolutional neural networks (CNNs) [16-22]. Classical implementations rely on structured kernel operations applied over local patches, with computational complexity scaling linearly with the number of positions and filter weights [17]. Although recent advances in quantum linear algebra have enabled partial solutions to matrix operations [18] using HHL [9], quantum singular value estimation (SVE) [19], and SWAP test techniques [26-29], these methods typically incur high circuit depth and do not efficiently exploit data sparsity [22] - an important property in real-world applications such as edge detection and compressed sensing [23].

Several prior studies have attempted to bridge convolution and quantum computation through diverse strategies. Kerenidis et al. (2019) introduced a quantum algorithm for convolutional neural networks (QCNNs) [16] using circulant matrix encoding, but their method required dense state preparation and lacked efficient patch handling. Gitiaux (2022) proposed a generalized multi-SWAP test protocol to enable parallelized inner product estimation, aligning with the core idea of our batch convolution framework [24]. Unlike these works, our method explicitly exploits sparsity in both image and kernel encoding, offering scalable feature extraction without relying on circulant structure or oracle assumptions.

In this work, we propose a quantum-efficient convolution framework that reformulates the classical convolution operation as a sparse matrix multiplication. We introduce a novel reshaping format that transforms convolution into a structured matrix-vector product between sparsely encoded image patches and doubly block-Toeplitz-filtered kernels. Our method employs key-value sparse state preparation and leverages the SWAP test to estimate inner products with low circuit depth. This design enables the efficient computation of convolution outputs in superposition, reducing sampling overhead and allowing scalability to larger image dimensions. Furthermore, we show how our formulation is amenable to batching and hybrid quantum-classical integration.

The remainder of this paper is organized as follows: Section 2 introduces the foundational background on SWAP test circuits, convolution matrix formulation, sparse quantum state preparation, and QRAM-based access methods. Section 3 presents our proposed algorithm, analyzes its complexity, and compares it against classical and prior quantum methods. Section 4 concludes the work and outlines future research directions in quantum-enhanced signal processing.

## 2 Classical and Quantum Foundations of Convolution: Theoretical Background and Problem Setting

This section presents the mathematical and quantum mechanical preliminaries necessary to support the formal development of our algorithm. We begin by establishing notation conventions and essential definitions from linear algebra, quantum information, and complexity theory. Just a few notational conventions before getting into more detail, $\mathbb{R}$, $\mathbb{N}$, $\mathbb{C}$, and $\mathbb{R}^+$ denotes the real numbers, integers, complex numbers, and positive real numbers. For algorithmic complexity, we adopt standard big-O notation, $O(f(n))$, which indicates a running time upper bounded by $cf(n)$ for a fixed $c \in \mathbb{R}^+$ and sufficiently large $n \in \mathbb{N}$. To suppress polylogarithmic factors that are often insignificant in asymptotic analysis, we also employ the soft-O notation, $\tilde{O}(\ )$, which represents $O(f(n)polylog(n))$.

For a matrix $A \in \mathbb{R}^{n \times n}$, we denote $A \in \mathbb{R}^{n \times n}$ as the number of nonzero entries in A. The set $[n] = \{1, 2, \ldots, n\}$ denotes the canonical index set of size n. The standard basis for $\mathbb{R}^n$ is represented as $\{e_1, e_2, \ldots, e_n\}$, where each $e_i$ is the unit vector with 1 in the i-th position and 0 elsewhere. The $l_p$ norm of a vector $x \in \mathbb{R}^n$ is defined as $|x|_p := \left(\sum_{i \in [n]} |x_i|^p\right)^{1/p}$ with the special case of the $l_\infty$-norm given by $|x|_\infty = max_{i \in [n]} |x_i|$.

### 2.1 The SWAP Test

A key subroutine in our proposed framework is the SWAP test, a well-established quantum circuit used to estimate the inner product between two quantum states [20, 25]. This technique plays a foundational role in quantum machine learning, quantum fingerprinting, and low-depth quantum matrix multiplication schemes. Consider two normalized quantum states $|\phi\rangle$ and $|\psi\rangle$ that encode classical vectors. The initial system is prepared in the composite state $|0\rangle \otimes |\phi\rangle \otimes |\psi\rangle$, where the first qubit acts as an ancillary control. The protocol proceeds as follows (Figure 1): First, a Hadamard gate is applied to the control qubit, yielding the intermediate state $|+\rangle \otimes |\phi\rangle \otimes |\psi\rangle$. A controlled-SWAP gate is applied to the second and third qubits, conditioned on the control qubit. This produces the entangled state $\frac{1}{\sqrt{2}}(|0\rangle \otimes |\phi\rangle \otimes |\psi\rangle + |1\rangle \otimes |\psi\rangle \otimes |\phi\rangle)$. Then, a second Hadamard gate is applied to the control qubit. After this step, the full state becomes:

$$\frac{1}{2}(|0\rangle|\phi\rangle|\psi\rangle + |1\rangle|\phi\rangle|\psi\rangle + |0\rangle|\psi\rangle|\phi\rangle - |1\rangle|\psi\rangle|\phi\rangle) =$$
$$= \frac{1}{2}|0\rangle(|\phi\rangle|\psi\rangle + |\psi\rangle|\phi\rangle) + \frac{1}{2}|1\rangle(|\phi\rangle|\psi\rangle - |\psi\rangle|\phi\rangle) \quad (1)$$

A projective measurement is then performed on the control qubit. The probability of measuring the state $|0\rangle$ is given by:

$$P(0) = \frac{1}{2}(\langle\phi|\langle\psi| + \langle\psi|\langle\phi|)\frac{1}{2}(|\phi\rangle|\psi\rangle + |\psi\rangle|\phi\rangle) = \frac{1 + |\langle\psi|\phi\rangle|^2}{2} \quad (2)$$

This probability encodes the squared magnitude of the inner product $\langle\psi|\phi\rangle$, which can be extracted by repeating the measurement a number of times and estimating the frequency of the $|0\rangle$ outcome. To achieve an additive ε-approximation of $\langle\psi|\phi\rangle^2$ with high confidence, the test must be repeated $O(1/\varepsilon^2)$ times. Importantly, each trial requires a state preparation of both $|\phi\rangle$ and $|\psi\rangle$, as quantum measurement collapses the entangled state and precludes reuse of the qubits. This necessitates efficient and repeatable state preparation, especially when applied in convolutional contexts where many inner products are computed in parallel. Despite this limitation, the SWAP test remains a powerful and low-depth approach to quantum inner product estimation. It offers significant advantages over alternative techniques such as phase estimation or HHL-based amplitude retrieval, especially in noisy intermediate-scale quantum (NISQ) regimes. Its circuit complexity is modest, involving only a small number of single- and two-qubit gates, making it amenable to implementation on near-term hardware.

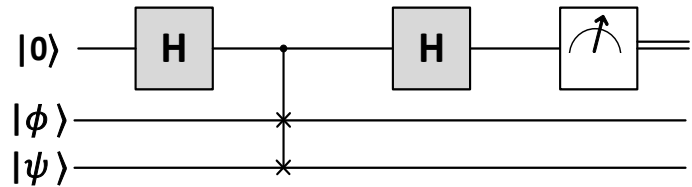

*Figure 1. The SWAP Test circuit.*

Originally introduced for quantum fingerprinting [20], the SWAP test has since been generalized to quantum machine learning, data classification, and most relevantly, matrix multiplication from quantum-to-classical or fully quantum data. When the desired precision is inverse polylogarithmic in input size—i.e., $O(1/poly\log(n))$, —then matrix multiplication using the SWAP test can be achieved in $\tilde{O}(n^2)$ time, representing a quadratic improvement over naïve classical methods.

## 2.2 Convolutional Layers

Convolutional operations form the computational backbone of modern image processing pipelines and play a central role in tasks such as edge detection, denoising, and hierarchical feature extraction. In classical optics, for example, the degradation of an image due to defocusing can be modeled as a convolution between the ideal sharp image and a kernel defined by the lens function—typically Gaussian in nature. This process, referred to in photography as bokeh, corresponds in computational image processing to blurring or smoothing, where convolution with a low-pass filter removes high-frequency components and preserves only coarse features [26].

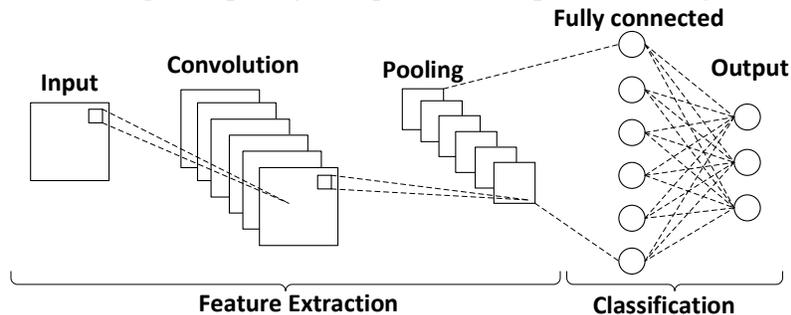

*Figure 2. The architecture of a traditional CNN.*

Beyond low-level vision tasks, convolution has become foundational in artificial intelligence, particularly in Convolutional Neural Networks (CNNs). CNNs are specialized deep learning architectures composed of cascaded convolution layers, often accompanied by non-linear activations, pooling layers, and normalization stages. In contrast to traditional fully connected architectures, CNNs rely on local connectivity and spatial weight sharing, enabling both parameter efficiency and translation invariance. Their dominance in classification, segmentation, and regression tasks stems from their ability to extract increasingly abstract representations of the input through repeated convolutional transformations (Figure 2) [27–29].

Each convolution layer operates over Input Feature Maps (Ifmaps) using a set of learnable filters or kernels. These kernels are typically represented as a four-dimensional tensor encompassing height, width, input channels, and output channels. When a filter is applied to a region of the Ifmaps, the resulting dot product captures the degree of similarity between the local input patch and the filter's learned pattern. The spatial scanning of the kernel over the input is known as the sliding window operation, and the set of outputs obtained forms a new representation called the Output Feature Map (Ofmap).

From a mathematical standpoint, each Ofmap channel corresponds to the convolution between a particular filter and one or more input channels. This operation acts as a feature extractor that highlights regions of the input image where a given feature—such as vertical edges, corners, or textures—is strongly present. The stronger the alignment between the filter weights and the local input pattern, the higher the activation value in the Ofmap. Figure 3 illustrates this process, where each kernel captures spatially localized information through repeated inner products, effectively compressing the image content into discriminative spatial features.

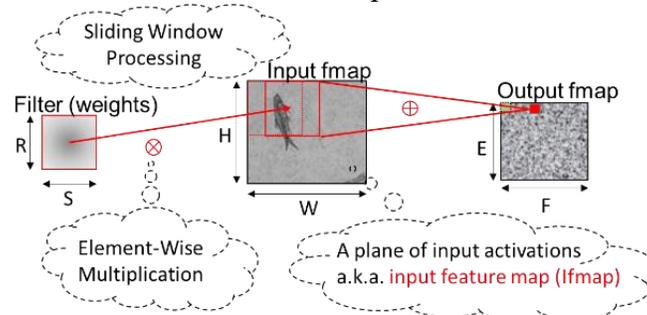

Figure 3. The convolving of one Ifmap channel by a single filter.

Thus, convolutional layers transform high-dimensional visual input into structured and compact representations suitable for downstream learning tasks. In the quantum setting, this classical convolutional mechanism must be reinterpreted in a form compatible with sparse state encoding and inner product evaluation via quantum circuits—topics explored in the following sections.

### 2.3 Vector State Preparation

A central challenge in the development of quantum algorithms for linear algebra, signal processing, and machine learning lies in the efficient preparation of quantum states from classical data. Specifically, given a real-valued vector $x \in \mathbb{R}^N$, one seeks to construct its normalized quantum encoding $|x\rangle = \frac{1}{|x|}\sum_{i \in [N]} x_i |i\rangle$, commonly referred to as vector state preparation. This transformation enables subsequent operations such as inner product estimation, quantum matrix-vector multiplication, and kernel evaluations to be performed in superposition.

The quantum advantage in this context arises from the logarithmic dimensionality of the Hilbert space: a vector in $\mathbb{R}^N$ can, in principle, be encoded using only $O(\log n)$ qubits. However, the efficiency of this encoding process is heavily dependent on the underlying memory model used to store and access classical data. In conventional classical memory architectures, data is accessed sequentially or via pointer-based random access, and preparing the vector state $|x\rangle$ from such a memory requires reading all coordinates—leading to a time complexity of $O(N)$ in the worst case, since each of the N elements must be visited explicitly.

To achieve exponential compression not only in storage but also in runtime, the encoding process must support querying vector entries in quantum superposition. This is the motivation behind the Quantum Random Access Memory (QRAM) model, which allows for coherent superpositions of classical addresses to be queried in parallel. Under this model, a vector x stored in QRAM can be accessed via an oracle such that $O_x: |i\rangle|0\rangle \rightarrow |i\rangle|x_i\rangle$, enabling the efficient construction of $|x\rangle$ using quantum circuits conditioned on amplitude encodings.

Nonetheless, fundamental lower bounds limit the performance of QRAM-based algorithms. A seminal result by Bennett et al. (1997) [30] established that the query complexity for state preparation under the oracle QRAM model is lower-bounded by $O(\sqrt{n})$ in the worst case. This result rules out the possibility of achieving general-purpose $O(polylog(n))$-time state preparation using standard QRAM oracles, thus motivating the need for refined models or data assumptions.

To overcome these limitations, more expressive memory models have been proposed. Notably, the augmented QRAM framework introduced by Kerenidis and Prakash (2014) [31] enables constant-time vector state preparation under structured access patterns and bounded sparsity. Their model assumes a preprocessing step that stores both values and cumulative norms, allowing for recursive sampling and coherent amplitude loading with logarithmic or even constant depth circuits. This augmented QRAM model underpins several quantum machine learning algorithms and provides a theoretical foundation for our proposed encoding strategy.

In our approach, we exploit the principles of augmented QRAM while avoiding costly preprocessing steps by introducing a sparse reshaping format tailored for convolutional data. This structure-aware encoding significantly reduces the effective query complexity and is well-suited for scenarios where input patches and kernels exhibit localized support. The ability to leverage this model for fast, parallelizable state preparation is one of the critical enabling features of our proposed quantum convolution framework.

The oracle Quantum Random Access Memory (QRAM) model has emerged as a foundational component in quantum algorithms that require coherent access to classical data. It serves as the standard abstraction for quantum query complexity studies and underlies algorithms such as Grover's search [2], amplitude amplification [14], and numerous quantum machine learning primitives [34]. At its core, QRAM enables queries over classical memory indices in coherent superposition, allowing the quantum circuit to perform data-dependent transformations across exponentially many states in parallel. Formally, if a dataset $\{x_i\}_{i \in [N]}$ is stored across N memory cells of QRAM, the oracle model supports the following transformation effectively enabling quantum parallel access to data. This structure is essential for quantum inner product estimation, matrix-vector multiplication, and amplitude encoding.

$$\sum_{i \in [N]} \alpha_i |i\rangle \rightarrow \sum_{i \in [N]} \alpha_i |i, x_i\rangle \quad (3)$$

Despite its theoretical utility, physically realizing a scalable and low-latency QRAM device remains a substantial challenge. The naive method of converting a classical RAM to a quantum-compatible system requires constructing quantum superpositions over all memory addresses, with depth scaling as $O(\log N)$. This conversion, while efficient in principle, imposes an exponentially large quantum coherence burden that limits its near-term practicality.

To mitigate these limitations, multiple architectural proposals for QRAM have been introduced. One prominent model is the bucket brigade architecture, proposed by Giovannetti, Lloyd, and Maccone [36]. In this design, the quantum control signals are routed through a binary tree of quantum switches, which reduces the number of active quantum gates at any time step and thereby lowers the resource footprint. The query time in this architecture scales as $O(log^2 N)$, trading spatial parallelism for circuit depth and fault tolerance. Other architectural variants, including fan-out trees and hybrid classical-quantum control schemes, have been explored in the literature [32, 33], each offering trade-offs between scalability, noise resilience, and gate depth.

Importantly, the oracle QRAM is a read-only memory model. The address register is used solely to identify the memory location, and it does not interact directly with the data contents in a non-linear fashion. As a result, transformations of the form $|i\rangle \rightarrow |i \oplus x_i\rangle$ are not achievable within the standard QRAM abstraction. This limitation is non-trivial in cryptographic and reversible logic applications where entanglement between index and content is required.

Furthermore, the unitary nature of quantum memory access constrains the possible transformations and enforces reversibility, which complicates QRAM circuit design compared to classical memory fetch operations. While theoretical constructions demonstrate asymptotically optimal access times (e.g., $O(\log N)$), their gate complexity and error rates are significant obstacles for hardware realization.

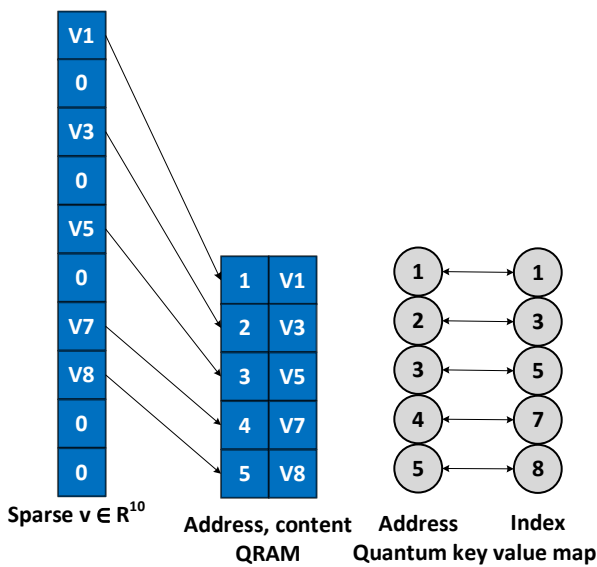

*Figure 4. Sparse vector state preparation using a quantum key-value map [31].*

Nevertheless, given the crucial role of QRAM in enabling scalable quantum algorithms, especially those involving structured datasets such as images or signals, continued development of feasible QRAM implementations remains a key enabler of quantum advantage. In this work, we leverage QRAM-inspired data models to propose a sparsity-aware convolution encoding scheme, which reduces the effective size of required state preparation while remaining consistent with realistic memory access models.

While general-purpose quantum state preparation of classical vectors remains a bottleneck due to its linear query complexity, the problem can be significantly alleviated in cases where the vector is sparse—that is, when the number of non-zero elements $nnz(x) \ll N$. Classical-to-quantum data loading typically incurs a cost of $O(N)$ when employing amplitude amplification techniques [14], matching the lower bounds established for unstructured search problems such as Grover's algorithm [2]. This runtime is considered optimal in the absence of additional structure or pre-processing assumptions.

In the context of sparse vectors, however, this complexity can be reduced further. Specifically, for a vector $x \in \mathbb{R}^n$ with $nnz(x)$ non-zero elements, the quantum state $|x\rangle$ can be prepared in time $O(\sqrt{nnz(x)})$ using a quantum key-value map structure introduced in [31], in combination with amplitude amplification. This key-value map provides a quantum-accessible mechanism for storing the positions of non-zero elements in an auxiliary address space, enabling efficient sparse indexing.

To illustrate, let $v \in \mathbb{R}^n$ be a sparse vector whose non-zero entries $v_{t_i}$, for cxxx $i \in nnz(v)$, are stored in contiguous memory locations indexed by $i$. Their reduction is illustrated in Figure 4. A key-value mapping $(K_i, V_i)$ is maintained such that $K_i = i$ and $V_i = t_i$, where $t_i$ is the original index in the full vector space. This map can be queried in superposition:

$$\sum_{i \in [n]} \alpha_i |K_i\rangle \leftrightarrow \sum_{i \in [n]} \alpha_i |V_i\rangle \qquad (4)$$

enabling indirect addressing of non-zero values.

The preparation proceeds in two phases: i. initial Sparse Superposition, a state $|v'\rangle == \sum_{i \in [nnz(v)]} v_{t_i} |i\rangle$ is constructed using amplitude amplification, in time $O(\sqrt{nnz(v)})$, ii. index Remapping via Key-Value Map; The quantum key-value oracle is applied to convert the intermediate state into $|v\rangle = \sum_{i \in [nnz(v)]} v_{t_i} |t_i\rangle$ corresponding to the original encoding of the sparse vector. The quantum key-value oracle can be initialized classically in time $O(nnz(v))$, and its query can be executed in time $O(polylog(n))$. This reduction demonstrates that amortized sparse state preparation is achievable in near-constant time under structured access assumptions.

More generally, for a vector $x \in \mathbb{R}^n$, the authors in [31] show that $k$ independent copies of the state $|x\rangle$ can be generated in time $O(nnz(x) + k \cdot \sqrt{nnz(x)} \cdot |x|_\infty)$ when utilizing an augmented QRAM augmented with quantum key-value maps. This model assumes that queries return not only value but positional metadata, which supports recursive amplitude loading with minimal overhead.

Additionally, the augmented QRAM architecture proposed in [31] enables constant-time vector state preparation when vectors are pre-encoded into logarithmic-depth memory trees. Given a vector $v \in \mathbb{R}^n$ and $l = \lceil \log n \rceil$, the mapping expressed in Eq. 5 can be implemented the unitary operator $|v\rangle = \frac{1}{\sqrt{|v|}} \sum_{i \in [n]} v_i |i\rangle$ in time $O(polylog(n))$, assuming an offline preprocessing stage. This sparse-access quantum memory strategy forms a key component of our proposed quantum convolution algorithm. By constraining the input and kernel tensors to exhibit localized support—typical in edge detection and feature extraction—we dramatically reduce the cost of quantum state preparation, enabling practical application of quantum circuits for high-dimensional signal processing.

$$|i, 0^l\rangle \rightarrow |i, v_i\rangle \qquad (5)$$

The accelerated performance of vector state preparation in quantum algorithms leveraging augmented QRAM is largely attributed to efficient pre-processing techniques that transform classical data into a form suitable for fast quantum access. In particular, the framework proposed in [31] introduces an optimized algorithm for inserting sparse vectors into an augmented QRAM architecture and for executing efficient queries thereafter, as specified by the unitary mapping in Eq. 5.

The procedure begins by pre-processing a vector $x \in \mathbb{R}^n$ prior to insertion. This step incurs an overhead of $O(nnz(x))$, where $nnz(x)$ denotes the number of non-zero elements in $x$. The preprocessing includes normalization of the vector and the construction of auxiliary metadata structures required for key-value access and state re-mapping. Notably, the vector x is transformed into a unit-norm state, ensuring that the resulting quantum encoding conforms to the amplitude encoding convention.

To enhance scalability, the insertion algorithm can be parallelized. Given access to $p$ parallel processing units, the pre-processing complexity can be reduced to $O(nnz(x/p))$, enabling substantial speedups when implemented on modern multi-core CPUs or GPU architectures. Each sub-vector is independently normalized and then stored into its respective segment of the QRAM, ensuring that the final vector can be assembled coherently during quantum queries.

Importantly, this process introduces minimal hardware overhead to the QRAM system. The augmented QRAM is designed to accommodate auxiliary metadata, such as the vector norms of stored entries, without increasing circuit depth significantly. These norms are maintained in a classical register $q \in \mathbb{R}^m$, where $m$ denotes the number of distinct vectors encoded in the memory. The use of such metadata allows for hybrid quantum-classical operations, particularly beneficial for applications requiring repeated state preparation or statistical sampling over input distributions.

*Table 1. Comparison of vector state preparation algorithms for creating C copies of $|x\rangle$ for $x \in \mathbb{R}^N$.*

| Method | Running Time | Extra Resources |
|---|---|---|
| Amplitude Amplifications [14] | $\tilde{O}(C\sqrt{N}|x|_\infty)$ | None |
| Sparsity-Aware Amplitude Amplifications [31] | $\tilde{O}(nnz(x) + C\sqrt{nnz(x)}|x|_\infty)$ | Key Value Map |
| Augmented QRAM [31] | $\tilde{O}(nnz(x) + C)$ | Metadata and Key Value Map |
| Parallel Augmented QRAM [31] | $\tilde{O}(nnz(x)/p + C)$ | A Classic Computer with $p$ Parallel Processing Units |

The time complexity for preparing $C$ copies of a normalized vector state using this framework can thus be expressed in two components: (i) the pre-processing time to prepare and insert the classical vector, and (ii) the state generation time per copy. This modular breakdown is presented in Table 1, which summarizes the computational complexity and resource requirements of several state preparation techniques, including those employing sparse data structures, amplitude amplification, and key-value mapping.

Overall, this approach enables practical and scalable integration of classical datasets into quantum algorithms by balancing QRAM memory model assumptions with efficient classical pre-processing and parallelism.

### 2.4 Convolution Operation

Convolutional operations lie at the heart of modern image processing and deep learning frameworks, where they are employed to extract hierarchical spatial features from structured input data. In the classical setting, convolution can be formalized as a sliding inner product between local regions of an input tensor and a corresponding set of filters.

Consider the simplified scenario illustrated in Figure 5(a), where we perform a single convolution operation over an input tensor consisting of $C$ channels, each of size $\mathbb{R}^{H \times W}$. Let there be a corresponding set of $C$ filters (or kernels), each of size $\mathbb{R}^{R \times S}$, such that the number of filters matches the number of channels. The convolution proceeds by sliding each filter across its corresponding channel and computing the elementwise inner product over each local window. The resulting scalar values from each of the $C$ filter–channel pairs are then aggregated—typically via summation—into a single output scalar for that window position.

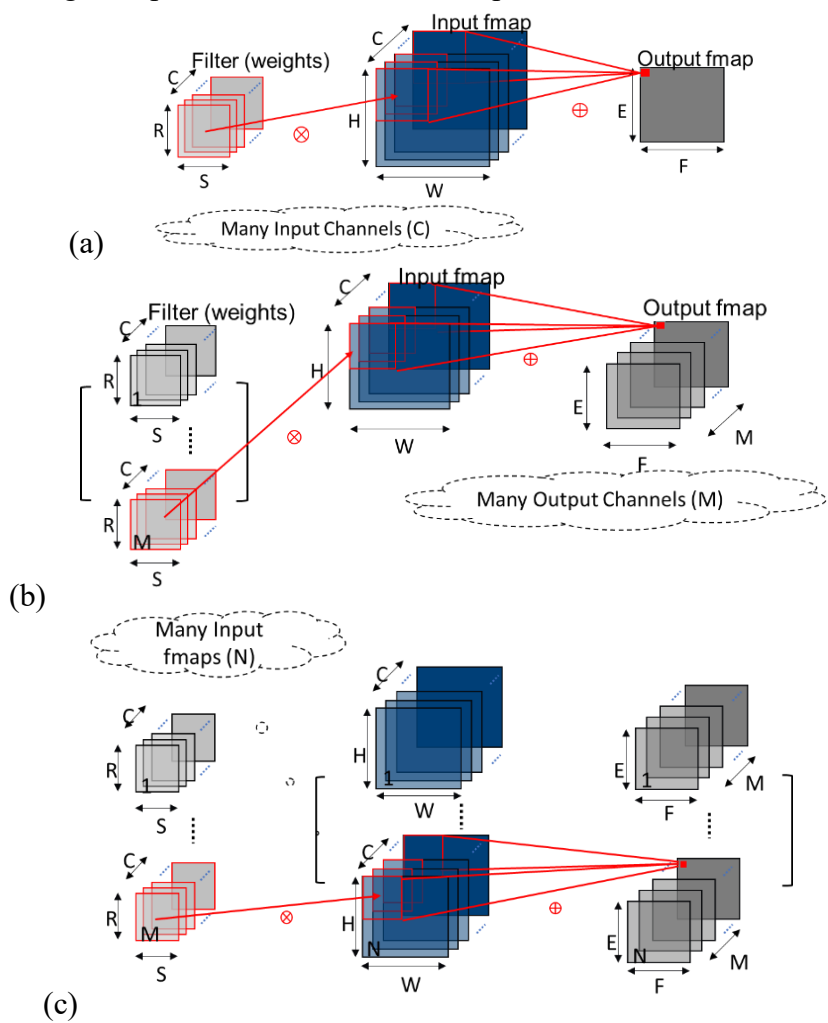

*Figure 5. (a) The convolution of a 3-D input tensor by one 3-D kernel tensor. (b) The several convolutions of a 3-D input tensor by one 4-D kernel tensor. (c) The several convolutions of a batch of inputs by one 4-D kernel tensor.*

This operation yields a 2D output tensor of shape $\mathbb{R}^{E \times F}$, where the output spatial dimensions are determined by the following expressions (assuming unit stride and no padding for clarity):
$$\begin{cases} E = H - R + 1 \\ F = W - S + 1 \end{cases} \tag{6}$$

To formalize the data structure, observe that a tensor is a natural generalization of a matrix to higher dimensions. A grayscale image is represented as a 2D matrix in $\mathbb{R}^{H \times W}$, where each element corresponds to an 8-bit intensity value. RGB images introduce an additional channel axis, forming a 3D tensor $\mathbb{R}^{H\times W\times 3}$. More generally, image batches used in convolutional neural networks (CNNs) are represented as 4D tensors with dimensions $\mathbb{R}^{N\times H\times W\times C}$, where $N$ denotes the batch size and $C$ the number of channels.

While padding and stride are essential parameters in practical convolutional architectures—modifying how filter windows are applied and determining output dimensions—we deliberately omit them in this formulation to maintain focus on the core algorithmic structure. These parameters can be seamlessly integrated into our proposed quantum convolution algorithm if needed.

The generalized convolution problem, as shown in Figure 5(b), involves applying multiple filter sets to a single input image to extract diverse features. Given an input tensor $\mathbb{R}^{H\times W\times C}$ and a bank of $M$ distinct filters, each filter itself is a 3D tensor of shape $\mathbb{R}^{R\times S\times C}$, capturing spatial and channel-wise correlations. Collectively, the entire set of filters is represented as a 4D tensor $\mathbb{R}^{R\times S\times C\times M}$.

The output tensor $Y \in \mathbb{R}^{E\times F\times M}$ aggregates the responses from all $M$ filters across all spatial locations. Each scalar element $Y_{EFM}$ of the output can be computed via a three-way summation:

$$-Y_{iE,jF,dM} = \sum_{i=0}^{R-1}\sum_{j=0}^{S-1}\sum_{k=0}^{C-1} X_{iE+i,jF+j,k} \times K_{i,j,k,dM} \tag{7}$$

as shown in Eq. 7. This formulation precisely captures the weighted accumulation of local patches across channels and filters, central to convolutional feature extraction.

To further exploit computational parallelism, especially in high-throughput applications, input images are often batched. A batch size $N$ introduces another axis to the input tensor, resulting in $X \in \mathbb{R}^{N\times H\times W\times C}$, with corresponding output $Y \in \mathbb{R}^{N\times E\times F\times M}$, as illustrated in Figure 5(c).

For a detailed mathematical treatment of multidimensional convolution operations and their practical implementation across neural architectures, we refer readers to [34]. Additionally, conventional image processing applications such as edge detection with Canny filters [35] are special cases of this generalized convolution structure, differing primarily in kernel design and output interpretation.

## 3 Quantum Matrix Multiplication for Convolution: Methods and Analysis

### 3.1 The Reshaping Method

The core objective of this study is to reformulate the convolution operation defined in Eq. 7 as an efficient matrix multiplication problem, thereby enabling its acceleration via quantum linear algebra primitives. This reformulation is grounded in the observation that convolution operations can be expressed as generalized matrix multiplications over suitably reshaped inputs and kernels. Previous work, such as that by Chen et al. [16], explored this approach by mapping the convolution operation to matrix multiplication using Toeplitz and circulant matrices [36, 37]. Their method constructs an intermediate matrix $\tilde{X}$, which is structured such that each row encodes a vectorized local patch of the input tensor $X$, consistent with the sliding window mechanism of standard convolution.

However, as depicted in Figure 6(a), their construction incurs substantial redundancy in the form of duplicated nonzero elements and requires storing pixel positions and intensities explicitly in quantum-accessible data structures. This reshaping process not only introduces high computational overhead but also complicates the propagation of feature maps across multiple layers of a deep convolutional architecture, necessitating reformatting of $Y$ to be compatible with subsequent layers.

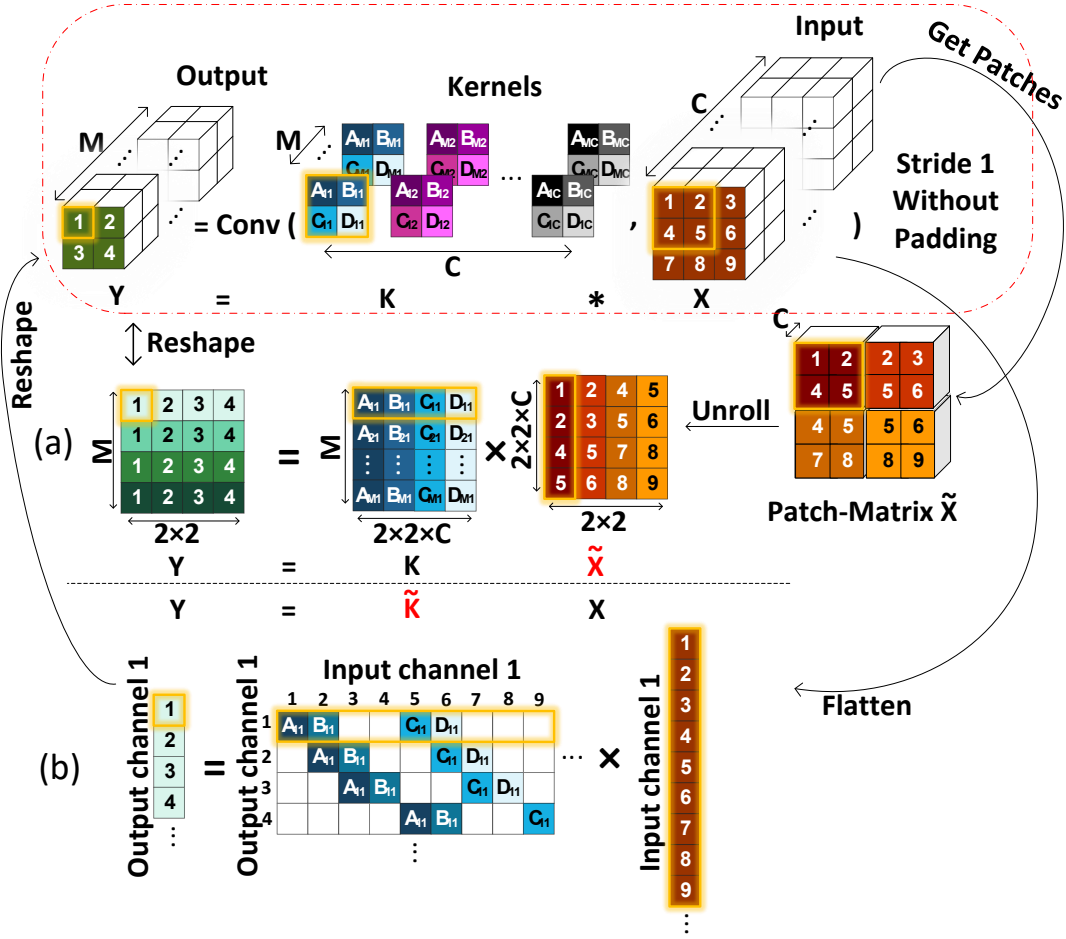

*Figure 6. The basic idea of the reshaping method based on (a) Toeplitz and (b) Doubly Block-Toeplitz matrices considering R=S=2 and H=W=3 and Stride=1 and without Padding.*

To address these limitations, we propose an alternative quantum-friendly reshaping strategy that significantly reduces preprocessing complexity. This approach is based on the doubly block-Toeplitz (DBT) representation of convolutional filters, which preserves the spatial integrity of the input tensor $X$ while enabling a direct matrix formulation of the convolution operation. As shown in Figure 6(b), our method avoids expanding $X$ into a redundant patch matrix and instead keeps it intact while reshaping the convolutional kernel tensor $K$ into a structured 2D matrix $\widetilde{K}$.

Formally, let the input tensor $X \in \mathbb{R}^{H \times W \times C}$ be flattened into a column vector of dimension HWC, while the filter tensor $K \in \mathbb{R}^{R \times S \times C \times M}$ is reshaped into a matrix $\widetilde{K} \in \mathbb{R}^{EFM \times HWC}$, where each row of $\widetilde{K}$ encodes a kernel window aligned with a valid spatial region of $X$. The inner product between each such row and the flattened $X$ yields a scalar value in the output vector $Y \in \mathbb{R}^{E \times F \times M}$, thus transforming the convolution into the following matrix operation:

$$\widetilde{K} \bullet X = Y \tag{8}$$

as presented in Eq. 8.

Each element $Y_p$ of the output corresponds to the inner product between the p-th row of $\widetilde{K}$ and the flattened input vector $X$, where $p \in [EFM]$. The output vector can subsequently be reshaped into a 3D tensor of dimensions $E \times F \times M$, where each $EF$-block corresponds to the feature map generated by one of the $M$ filters. Importantly, the structure of $\widetilde{K}$ allows for a highly sparse and structured representation, particularly amenable to efficient loading into quantum memory using sparse QRAM techniques (as discussed in Section 2). This facilitates the use of quantum inner product estimation algorithms, such as the SWAP test, to compute the convolution results in quantum parallelism with reduced circuit depth.

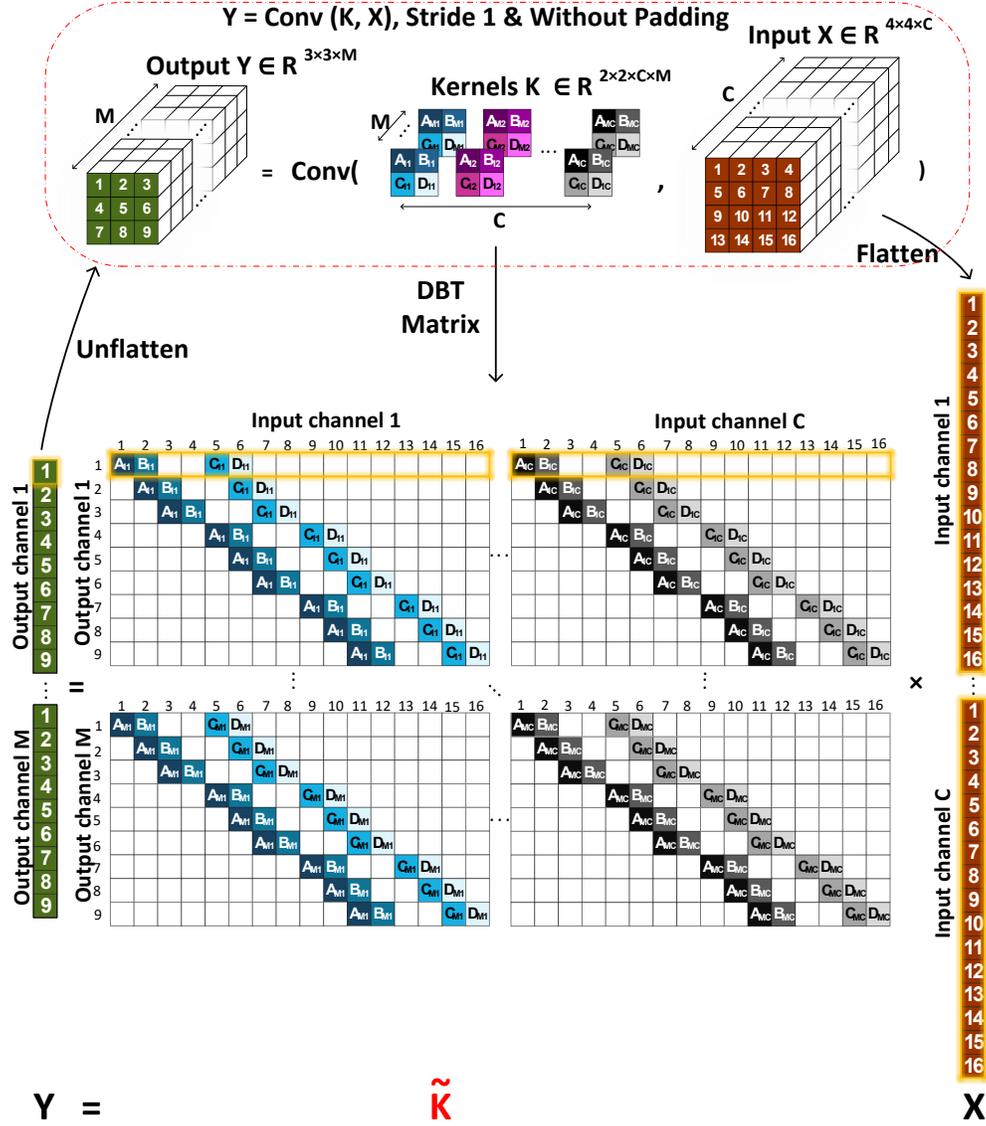

Figure 7. The complete view of the proposed pattern.

To enable batch processing, especially relevant for large-scale learning tasks, the approach is naturally extended to accommodate multiple inputs. Let $N$ denote the batch size. Each input tensor $K(i) \in \mathbb{R}^{H \times W \times C}$ is vectorized and stacked column-wise to form a matrix $X \in \mathbb{R}^{HWC \times N}$. The convolution output for the entire batch is then obtained via the matrix multiplication $\widetilde{K}X = Y$ resulting in an output matrix $Y \in \mathbb{R}^{EFM \times N}$, where each column corresponds to the flattened output tensor for an input in the batch. This structure is inherently compatible with quantum data loading and processing pipelines, enabling efficient quantum acceleration for convolutional operations.

In summary, our proposed reshaping framework circumvents the redundancy and complexity of earlier Toeplitz-based constructions and offers a scalable, sparse, and structured format tailored for efficient integration with low-depth quantum circuits. The reduction in preprocessing cost and the alignment with quantum matrix multiplication protocols make this approach a promising candidate for near-term quantum machine learning architectures.

### 3.2 The Quantum Algorithm

Building upon the sparse matrix reshaping strategy introduced in Section 3.1, we now develop a quantum algorithm for performing the core computation: the matrix multiplication between the reshaped filter matrix $\widetilde{K} \in \mathbb{R}^{EFM \times HWC}$ and the batch-encoded input matrix $X \in \mathbb{R}^{HWC \times N}$. This operation is executed through the estimation of inner products between corresponding rows of $\widetilde{K}$ and columns of $X$, using quantum amplitude-based protocols. To this end, we define quantum states proportional to the normalized rows of the kernel matrix and the normalized columns of the input matrix as described in Eq. 9 and 10.

$$|K_p\rangle = \frac{1}{\|K_p\|} \sum_{r=0}^{HWC-1} K_{pr} |r\rangle \text{ for } p \in [EFM]. \tag{9}$$

$$|X_q\rangle = \frac{1}{\|X_q\|} \sum_{s=0}^{HWC-1} X_{sq} |s\rangle \text{ for } q \in [N]. \tag{10}$$

Assuming these classical vectors are preprocessed and stored in quantum-accessible memory (e.g., augmented QRAM), the corresponding quantum states $|K_p\rangle$ and $|X_q\rangle$ can be efficiently loaded via queries defined in Eq. 11.

$$\begin{cases} |p\rangle|0\rangle \mapsto |p\rangle|K_p\rangle \\ |q\rangle|0\rangle \mapsto |q\rangle|X_q\rangle \end{cases} \tag{11}$$

These state preparation routines rely on data structures that enable superposition access, and their complexities are polylogarithmic in the ambient vector dimension and sublinear in the number of nonzero entries, as summarized in Table 1 and discussed extensively in Section 2.

To estimate the inner product $\langle K_p|X_q\rangle$, we adapt the generalized swap-test-based circuit proposed in [38], which computes $|\langle\psi|\phi\rangle|^2$ with logarithmic-depth circuits and minimal ancilla overhead. The procedure begins by constructing the controlled state preparation, followed by initializing a quantum register in the state $|\psi_0\rangle$ and applying a Hadamard gate to the third (ancilla) qubit and a controlled preparation of $|K_p\rangle$ and $|X_q\rangle$, yielding state expressed in Eq. 12.

$$|\psi_0\rangle = |p\rangle|q\rangle|0\rangle|0\rangle \mapsto \frac{1}{\sqrt{2}}(|p\rangle|q\rangle|0\rangle|0\rangle + |p\rangle|q\rangle|1\rangle|0\rangle) \tag{12}$$

$$\mapsto |\psi_1\rangle \frac{1}{\sqrt{2}}(|p\rangle|q\rangle|0\rangle|K_p\rangle + |p\rangle|q\rangle|1\rangle|X_q\rangle)$$

Now, applying a second Hadamard gate to the third register yields the final quantum state:

$$|\psi_2\rangle = \frac{1}{2}|p\rangle|q\rangle(|0\rangle(|K_p\rangle + |X_q\rangle) + |1\rangle(|K_p\rangle - |X_q\rangle)) \tag{13}$$

The measurement probability of observing the ancilla qubit in the state $|0\rangle$ is given by:

$$P_{pq}(0) = \frac{1}{4}(2 + 2\langle K_p|X_q\rangle) = \frac{1 + \langle K_p|X_q\rangle}{2} \tag{14}$$

This implies that by repeating the procedure a sufficient number of times and recording the measurement outcomes, one can efficiently estimate the real part of the inner product $\langle K_p|X_q\rangle$ to the desired precision.

Finally, the quantum state $|\psi_2\rangle$ can be compactly expressed as:

$$|\psi_2\rangle = |p\rangle|q\rangle(\sqrt{P_{pq}}|0, y_{pq}\rangle + \sqrt{1 - P_{pq}}|1, y'_{pq}\rangle) \tag{15}$$

where $|y_{pq}\rangle$ and $|y'_{pq}\rangle$ denote auxiliary work registers conditioned on the measurement outcome of the third qubit.

This framework enables efficient quantum estimation of each entry $Y_{p,q} = \langle K_p|X_q\rangle$, forming the complete convolution output tensor $Y \in \mathbb{R}^{EFM \times N}$. In contrast to classical evaluation, the quantum implementation provides a logarithmic circuit depth (neglecting query time) and supports amplitude-based sampling, making it suitable for NISQ-era quantum processors and fault-tolerant platforms alike.

Given that the reshaped kernel matrix $\widetilde{K} \in \mathbb{R}^{EFM \times HWC}$ has $EFM$ rows and the batched input matrix $X \in \mathbb{R}^{HWC \times N}$ has $N$ columns, the inner product estimation routine described previously can be executed in superposition over all $(p, q) \in [EFM] \times [N]$. This allows for efficient parallel estimation of all output elements in the convolution result tensor $Y \in \mathbb{R}^{EFM \times N}$. By preparing the registers in uniform superposition and performing the controlled inner product estimation circuit, the system evolves to the entangled state:

$$\frac{1}{\sqrt{H \times W \times C}} \sum_p \sum_q |p\rangle|q\rangle(\sqrt{P_{pq}}|0, y_{pq}\rangle + \sqrt{1 - P_{pq}}|1, y'_{pq}\rangle) \tag{16}$$

From this state, the probability of observing the triplet $(p, q, 0)$ in the first three registers is given by:

$$P_0(p, q) = \frac{P_{pq}}{H \times W \times C} = \frac{1 + \langle K_p | X_q \rangle}{2(H \times W \times C)} \tag{17}$$

which is directly related to the inner product between the p-th kernel row and the q-th input column.

It is important to note that the value $\langle K_p|X_q\rangle$ corresponds to the normalized inner product between the respective vectors. The actual inner product $(K_p, X_q)$ is obtained via: $(K_p, X_q) = \|K_p\|\|X_q\|\langle K_p|X_q\rangle$ where the norms $\|K_p\|$ and $\|X_q\|$ are known from metadata stored in the augmented QRAM during state preparation (see Section 2). This probabilistic sampling framework ensures that positions $(p, q)$ corresponding to large values in the convolution output $Y$ are observed with higher probability. As a consequence, the quantum algorithm naturally prioritizes the extraction of semantically significant features from the input image - such as edges or textures - as they produce larger inner products and thus dominate the probability distribution.

From an information-theoretic perspective, this implies that a small number of measurements can be sufficient to recover the dominant components of $Y$, enabling feature extraction with reduced sampling complexity. The output values indicate how strongly each kernel feature is activated across the spatial regions of the input. Low output values imply poor match to the filter pattern, while high values signify close alignment with salient features.

Nonetheless, to reconstruct the full output tensor $Y$ with a desired precision, the state preparation and measurement process must be repeated multiple times. The required number of repetitions depends on the approximation threshold for the inner product values, which in turn governs the granularity of abstraction extracted from the input batch.

### 3.3 Analysis and Comparison

As emphasized in prior sections, the proposed sparse reshaping pattern offers a significant advantage over previous methods such as that of [16]. Notably, our formulation eliminates the need for complex pixel manipulation or transformation of the input image data. Instead, each input tensor is simply flattened into a one-dimensional vector, a process that can be executed in constant time $O(1)$. This marks a sharp departure from the prior approach in [16], which required extensive preparation and patchification of the input tensor into high-dimensional Toeplitz matrices, incurring substantial redundancy and time complexity.

An important feature of our method is that the entire kernel reshaping process from a four-dimensional filter tensor $K \in \mathbb{R}^{RSCM}$ to the matrix $\widetilde{K} \in \mathbb{R}^{EFM \times HWC}$ is performed entirely classically in $O(R \times S \times C \times M)$ time. This classical pre-processing step is decoupled from the quantum circuit and is executed only once per kernel configuration, as opposed to the per-input cost incurred in [16], where each new input necessitates construction of an entirely new Toeplitz structure.

Table 2. *Complexity and practicality comparison between classical, Toeplitz-based, swap-test-based, and our sparse reshaping-based convolution approaches. (QMM: Quantum Matrix Multiplication)*

| Method | QRAM Complexity | Circuit depth | Preprocessing Time | State Prep Complexity | Suitability for NISQ |
|---|---|---|---|---|---|
| This work | $\tilde{O}(\sqrt{nnz(x)})$ | $\tilde{O}(1)$ | $O(nnz(x))$ | Efficient for sparse data | High |
| Toeplitz + QMM [16] | $\tilde{O}(n^2)$ | $O(n)$ | $O(n^2)$ | Dense QRAM encoding | Low |
| Swap Test [17] | $\tilde{O}(n)$ | $O(n)$ | $O(n)$ | Repetitive ancilla prep | Medium |

Moreover, this architectural choice aligns well with practical inference scenarios in edge devices and data streaming contexts, where a large volume of input samples is processed continuously against a relatively small set of fixed filters. In such cases, the amortized cost of quantum state preparation per input is dramatically reduced. The flattened inputs exhibit no redundancy and contain only the true data values, while the structured filter matrix $\widetilde{K}$- padded as necessary with zeros - introduces no additional cost in the quantum pipeline, as quantum state preparation complexity scales with the number of non-zero elements only.

Importantly, the presence of zero-padding in the kernel reshaping does not degrade algorithmic efficiency. Since quantum vector state preparation algorithms, including those leveraging amplitude amplification and augmented QRAM [35], are sensitive only to the sparsity structure (i.e., $nnz(v)$), the inclusion of structured zeros in $\widetilde{K}$ has no detrimental effect on query or preparation complexity. Additionally, our formulation permits straightforward incorporation of stride and padding hyperparameters into the kernel reshaping rules, enabling broad applicability to real-world convolutional architectures. The flexibility of this pattern supports variable image and kernel dimensions, batch processing strategies, and downstream quantum neural network architectures.

Resource planning in our approach is equally tractable. The number of qubits required for the algorithm depends logarithmically on the input size and linearly on the batch size, i.e., $O(logHWC + logN)$. This facilitates pre-allocation and efficient scheduling based on quantum hardware limitations.

Finally, although the swap-test-based estimation procedure requires re-preparing quantum states for each measurement round, this process continues only until the target precision or abstraction level is achieved. Hence, with a shallow and hardware-friendly circuit architecture, our algorithm supports parallel quantum convolution across large input batches and filter banks, offering tunable trade-offs between speed, fidelity, and resource usage.

The current method implements quantum inner product estimation between input and filter encodings using the SWAP test. While effective for static filters, this approach is not learnable or adaptive — a key limitation for machine learning tasks.

To extend the convolution operation for tasks like classification, regression, or feature learning, we propose enhancing the framework with Variational Quantum Circuits (VQCs) that learn optimal convolutional filters directly from data in a hybrid classical–quantum loop. A significant advancement of the proposed method is its potential integration with quantum machine learning (QML) frameworks. Specifically, we propose replacing fixed convolutional filters with parameterized quantum circuits, also known as Variational Quantum Circuits (VQCs). In this setting, the convolution kernel is not statically loaded but instead represented as a parameterized quantum state, $|\theta\rangle$, where $\theta$ denotes the set of trainable gate parameters. The convolution becomes a learned quantum operation: $y_{pq} = \langle \phi(X_q)|U(\theta_p)|0\rangle$ where $U(\theta_p)$ is a variational ansatz acting on a reference state and $\varphi(X_q)$ is the quantum encoding of the input patch or feature vector.

A classical optimizer evaluates the measurement outcomes (e.g., expectation value of a Pauli observable or a classification loss) and updates θ using gradient-based or gradient-free methods. This hybrid optimization loop allows the quantum convolution layer to adaptively learn the optimal filters — an essential characteristic for tasks such as classification or unsupervised feature learning.

We envision a quantum convolutional network (QCN) architecture where each variational layer replaces a classical convolution layer. Furthermore, this model can be fine-tuned using quantum-aware gradient descent techniques, including parameter shift rules or stochastic gradient descent with quantum evaluation.

To contextualize the performance and practical relevance of our proposed approach, we compare it with other quantum and classical convolution strategies in terms of key resource requirements. These include QRAM query complexity, circuit depth, preprocessing overhead, and overall suitability for NISQ-era quantum devices. As shown in Table 2, our method—based on sparse matrix encoding and swap-test-based inner product estimation—offers favorable asymptotic scaling for sparse inputs and avoids the deep circuits and dense QRAM encodings required by prior approaches such as the Toeplitz reshaping method and parallel swap-test schemes. This comparative assessment underscores the efficiency and hardware alignment of our architecture, particularly when processing streaming inputs in quantum convolution pipelines.

## 4 Conclusion

This work presents a quantum algorithmic framework for accelerating convolution operations—particularly those arising in image processing and feature extraction—by reformulating the classical convolution product as a quantum matrix multiplication problem. We demonstrated that the high computational cost of traditional convolutional layers can be substantially reduced by leveraging a sparse reshaping representation of both the input tensor X and the kernel tensor K, enabling their efficient encoding into quantum states.

The proposed methodology capitalizes on the observation that the convolution operation admits a compact matrix product formulation when the kernel tensor is reshaped using a doubly block-Toeplitz (DBT) pattern and the input is flattened into a single vector. In contrast to prior approaches, which incur significant overhead due to pixel-wise patchification and input redundancy, our scheme introduces only controlled sparsity in the filter matrix and entirely eliminates redundancy in the input representation. This sparse encoding not only reduces the number of non-zero elements but also minimizes the cost of state preparation and oracle queries, which are typically bottlenecks in quantum linear algebra routines.

By employing quantum inner product estimation—implemented via low-depth swap test circuits or amplitude-based interference—we estimate the output of the convolution layer without fully reconstructing the intermediate quantum states. The batch structure of the algorithm supports simultaneous convolution of multiple inputs with multiple filters, making the method well-suited to large-scale, high-throughput quantum inference pipelines.

Importantly, the algorithm is designed to be compatible with existing quantum memory architectures such as augmented QRAM, allowing for efficient loading of sparse vectors with polylogarithmic query times. The modular nature of the design ensures scalability, adaptability to arbitrary convolutional dimensions, and easy incorporation of stride and padding parameters.

In summary, this study advances the state of the art in quantum convolution by introducing a scalable, low-complexity algorithm grounded in sparse matrix reshaping and quantum-efficient computation. The proposed framework opens new avenues for hybrid quantum-classical neural network architectures and quantum-accelerated image processing applications, particularly in regimes where high-dimensional data and efficient inference intersect.

Future research can build on this work by exploring hardware-aware optimizations for reducing circuit depth and enhancing fidelity on noisy intermediate-scale quantum (NISQ) devices, particularly for large-batch convolution tasks. Integrating the proposed quantum convolution framework into hybrid quantum-classical machine learning models could also enable efficient quantum preprocessing layers in deep neural networks. Moreover, generalizing the reshaping pattern to support higher-dimensional convolutions (e.g., 3D for video or volumetric data) and adaptive or trainable quantum filters could open new possibilities for quantum-enhanced feature extraction. Finally, investigating error mitigation strategies and resource-efficient state preparation techniques remains crucial for realizing practical implementations on current quantum hardware.

## References

1. Kookani, A., Mafi, Y., Kazemikhah, P., Aghababa, H., Fouladi, K., Barati, M.: XpookyNet: advancement in quantum system analysis through convolutional neural networks for detection of entanglement. Quantum Mach. Intell. 6, 50 (2024). https://doi.org/10.1007/s42484-024-00183-y

2. Moghadam, A.H., Kazemikhah, P., Aghababa, H.: Using 1-Factorization from Graph Theory for Quantum Speedups on Clique Problems, https://arxiv.org/abs/2308.16827, (2023)

3. Ahmadkhaniha, A., Mafi, Y., Kazemikhah, P., Aghababa, H., Barati, M., Kolahdouz, M.: Performance Analysis of the Hardware-Efficient Quantum Search Algorithm. Int. J. Theor. Phys. 62, 168 (2023). https://doi.org/10.1007/s10773-023-05424-7

4. Shor, P.W.: Polynomial-Time Algorithms for Prime Factorization and Discrete Logarithms on a Quantum Computer. SIAM Rev. 41, 303–332 (1999). https://doi.org/10.1137/S0036144598347011

5. Shor, P.W.: Algorithms for quantum computation: discrete logarithms and factoring. In: Proceedings 35th Annual

Symposium on Foundations of Computer Science. pp. 124–134 (1994)

6. Grover, L.K.: A Fast Quantum Mechanical Algorithm for Database Search. In: Proceedings of the Twenty-Eighth Annual ACM Symposium on Theory of Computing. pp. 212–219. Association for Computing Machinery, New York, NY, USA (1996)

7. Long, G.L.: Grover algorithm with zero theoretical failure rate. Phys. Rev. A. 64, 22307 (2001). https://doi.org/10.1103/PhysRevA.64.022307

8. Toyama, F.M., van Dijk, W., Nogami, Y.: Quantum search with certainty based on modified Grover algorithms: optimum choice of parameters. Quantum Inf. Process. 12, 1897–1914 (2013). https://doi.org/10.1007/s11128-012-0498-0

9. Harrow, A.W., Hassidim, A., Lloyd, S.: Quantum Algorithm for Linear Systems of Equations. Phys. Rev. Lett. 103, 150502 (2009). https://doi.org/10.1103/PhysRevLett.103.150502

10. Clader, B.D., Jacobs, B.C., Sprouse, C.R.: Preconditioned Quantum Linear System Algorithm. Phys. Rev. Lett. 110, 250504 (2013). https://doi.org/10.1103/PhysRevLett.110.250504

11. Gall, F.: Quantum Algorithms for Matrix Multiplication. Presented at the (2013)

12. Le Gall, F., Nishimura, H.: Quantum Algorithms for Matrix Products over Semirings. In: Ravi, R. and Gørtz, I.L. (eds.) Algorithm Theory -- SWAT 2014. pp. 331–343. Springer International Publishing, Cham (2014)

13. A.Yu.Kitaev: Quantum measurements and the Abelian Stabilizer Problem, (1995)

14. Brassard, G., Høyer, P., Mosca, M., Tapp, A.: Quantum amplitude amplification and estimation. Quantum Comput. Inf. 53–74 (2002). https://doi.org/10.1090/conm/305/05215

15. Farhi, E., Goldstone, J., Gutmann, S., Lapan, J., Lundgren, A., Preda, D.: A Quantum Adiabatic Evolution Algorithm Applied to Random Instances of an NP-Complete Problem. Science (80-. ). 292, 472–475 (2001). https://doi.org/10.1126/science.1057726

16. Kerenidis, I., Landman, J., Prakash, A.: Quantum Algorithms for Deep Convolutional Neural Networks, (2019)

17. Shao, C.: Quantum Algorithms to Matrix Multiplication, (2018)

18. Aho, A. V, Hopcroft, J.E.: The design and analysis of computer algorithms. Pearson Education India (1974)

19. Kerenidis, I., Prakash, A.: Quantum Recommendation Systems. In: 8th Innovations in Theoretical Computer Science Conference. pp. 1–21 (2017)

20. Buhrman, H., Cleve, R., Watrous, J., de Wolf, R.: Quantum Fingerprinting. Phys. Rev. Lett. 87, 167902 (2001). https://doi.org/10.1103/PhysRevLett.87.167902

21. Garcia-Escartin, J.C., Chamorro-Posada, P.: swap test and Hong-Ou-Mandel effect are equivalent. Phys. Rev. A. 87, 52330 (2013). https://doi.org/10.1103/PhysRevA.87.052330

22. Buhrman, H., Špalek, R.: Quantum Verification of Matrix Products. In: Proceedings of the Seventeenth Annual ACM-SIAM Symposium on Discrete Algorithm. pp. 880–889. Society for Industrial and Applied Mathematics, USA (2006)

23. Zhang, X.-D., Zhang, X.-M., Xue, Z.-Y.: Quantum hyperparallel algorithm for matrix multiplication. Sci. Rep. 6, 24910 (2016). https://doi.org/10.1038/srep24910

24. Gitiaux, X., Morris, I., Emelianenko, M., Tian, M.: SWAP test for an arbitrary number of quantum states. Quantum Inf. Process. 21, (2022). https://doi.org/10.1007/s11128-022-03643-1

25. Li, P., Wang, B.: Quantum neural networks model based on swap test and phase estimation. Neural Networks. 130, 152–164 (2020)

26. Getreuer, P.: A Survey of Gaussian Convolution Algorithms. Image Process. Line. 3, 286–310 (2013)

27. LeCun, Y., Bengio, Y., Hinton, G.: Deep learning. Nature. 521, 436–444 (2015). https://doi.org/10.1038/nature14539

28. Russakovsky, O., Deng, J., Su, H., Krause, J., Satheesh, S., Ma, S., Huang, Z., Karpathy, A., Khosla, A., Bernstein, M., Berg, A.C., Fei-Fei, L.: ImageNet Large Scale Visual Recognition Challenge. Int. J. Comput. Vis. 115, 211–252 (2015).

https://doi.org/10.1007/s11263-015-0816-y

29. Simonyan, K., Zisserman, A.: Very Deep Convolutional Networks for Large-Scale Image Recognition, (2015)

30. Bennett, C.H., Bernstein, E., Brassard, G., Vazirani, U.: Strengths and Weaknesses of Quantum Computing. SIAM J. Comput. 26, 1510–1523 (1997). https://doi.org/10.1137/S0097539796300933

31. Prakash, A.: Quantum algorithms for linear algebra and machine learning., (2014)

32. Giovannetti, V., Lloyd, S., Maccone, L.: Quantum Random Access Memory. Phys. Rev. Lett. 100, 160501 (2008). https://doi.org/10.1103/PhysRevLett.100.160501

33. Giovannetti, V., Lloyd, S., Maccone, L.: Architectures for a quantum random access memory. Phys. Rev. A. 78, 52310 (2008). https://doi.org/10.1103/PhysRevA.78.052310

34. Wu, J.: Introduction to Convolutional Neural Networks. In: National Key Lab for Novel Software Technology. Nanjing University. China. p. 5. pp. 23, (2017)

35. Canny, J.: A computational approach to edge detection. Trans. pattern Anal. Mach. Intell. 679–698 (1986)

36. Gray, R.M.: Toeplitz and Circulant Matrices: A Review. Found. Trends® Commun. Inf. Theory. 2, 155–239 (2006). https://doi.org/10.1561/0100000006

37. Heinig, G., Bojanczyk, A.: Transformation techniques for Toeplitz and Toeplitz-plus-Hankel matrices II. Algorithms. Linear Algebra Appl. 278, 11–36 (1998). https://doi.org/https://doi.org/10.1016/S0024-3795(97)10043-X

38. Kerenidis, I., Landman, J., Luongo, A., Prakash, A.: q-means: A quantum algorithm for unsupervised machine learning, (2018)